# The re-entrant ferromagnetism due to symmetric exchange bias in two septuple-layer MnBi$_2$Te$_4$ epitaxial films

Dezhi Song[1], Fuyang Huang[1], Gang Yao[2], Haimin Zhang[1], Haiming Huang[1], Hang Yan[3], Jun Zhang[1*], Qinghua Zhang[4], Lin Gu[5], Xu-Cun Ma[3,6], Jin-Feng Jia[2,7,8], Qi-Kun Xue[4,6,7,8], Ye-Ping Jiang[1*]

1 Key Laboratory of Polar Materials and Devices, Department of Electronic, East China Normal University, Shanghai 200241, China.

2 Key Laboratory of Artificial Structures and Quantum Control (Ministry of Education), Tsung-Dao Lee Institute, School of Physics and Astronomy, Shanghai Jiao Tong University, Shanghai, 200240, China.

3 State Key Laboratory of Low-Dimensional Quantum Physics, Department of Physics, Tsinghua University, Beijing 100084, China.

4 Beijing National Laboratory for Condensed Matter Physics, Institute of Physics, Chinese Academy of Sciences, Beijing 100190, China.

5 Beijing National Center for Electron Microscopy and Laboratory of Advanced Materials, School of Materials Science and Engineering, Tsinghua University, Beijing 100084, China.

6 Frontier Science Center for Quantum Information, Beijing 100084, China.

7 Southern University of Science and Technology, Shenzhen 518055, China.

8 Quantum Science Center of Guangdong-HongKong-Macao Greater Bay Area, Shenzhen 518045, China.

*Corresponding author. Email: zhangjun@ee.ecnu.edu.cn, ypjiang@clpm.ecnu.edu.cn

MnBi$_2$Te$_4$ (MBT) is a typical magnetic topological insulator with an A-type antiferromagnetic (AFM) ground state. Here we prepared ultra-thin MBT films with controlled anti-site defects and observed rich doping-dependent magnetic behaviors. We find in one-septuple-layer MBT films a ferrimagnetic ground state and in two-septuple-layer ones a kind of re-entrant ferromagnetism (FM) that disappears with increased Bi-on-Mn doping in the FM Mn-layers. This re-entrant behavior is attributed to a kind of symmetric exchange-bias effect that arises in the presence of both AFM and FM sub-systems due to the introduction of high-dense Mn-on-Bi anti-site defects. Furthermore, all MBT films display spin-glass-like behaviors. Our work demonstrates rich magnetic behaviors originating from the competing magnetic interactions between Mn spins at different lattice positions in MBT.

In topological state of matters, the exchange interactions may change the band topology similarly to what spin-orbital interactions do to topological insulators (TIs), where in the latter case the spin-orbital field is momentum dependent. The additional introduction of exchange fields into TIs may make the band topology trivial or can also lead to the realization of the quantum spin Hall state of matter[1,2]. Compared with other magnetic doped TIs, MnBi$_2$Te$_4$ (MBT) stands for a family of intrinsic magnetic TIs[3-7], where the magnetism is inherent in their pure crystalline state.

MBT is composed of van-der-Waals (vdW) septuple layers (SL) of atoms stacked in the Te-Bi-Te-Mn-Bi-Te sequence (Fig. 1(a)). The MBT multilayers have an A-type anti-ferromagnetic (AFM) ground state[4,5,8], where the Mn spins align ferromagnetically inside each Mn-layer (with an out-of-plane easy axis) and anti-ferromagnetically between adjacent SLs. This kind of A-type magnetism makes MBT more interesting due to the prediction of the exotic thickness-dependent even-odd oscillation between the topological Chern insulators and axion insulators [6,9-13].

Experimentally, there is inevitable existence of anti-site defects in the Bi- and Mn-layers[14-17]. Here, by molecular beam epitaxy we achieved the tunable doping of Bi in the Mn layers (Bi$_{Mn}$), driving the ferromagnetic (FM) 1-SL MBT into a spin-glass (SG)-like state with heavy Bi$_{Mn}$ doping. In addition, in 2-SL MBT films, we observe the re-entrant FM behavior with the increasing temperature. We attribute this behavior to a kind of symmetric exchange-bias effect due to the simultaneous existence of FM and AFM sub-systems in 2-SL MBT. This re-entrant behavior disappears with the Bi$_{Mn}$ doping. A microscopic toy model is presented to account for these behaviors. Thus, the finely controlled doping of anti-site defects in MBT leads to the observation of rich doping-dependent magnetism that may be crucial to the topological properties of this kind of intrinsic magnetic TI.

In our MBT films prepared by the procedure described in the section I of [18], the dilute doping of Mn in Bi-layers (Mn$_{Bi}$) is nearly constant, while the Bi$_{Mn}$ doping in the Mn-layers is highly tunable. These result in the off-stoichiometric MBT in the form of Mn$_{1-x+2y}$Bi$_{2(1-y)+x}$Te$_4$, where Mn$_{Bi}$ doping is fixed at $y \approx 0.1$ (Fig. S1 of [18]) and Bi$_{Mn}$ doping $x$ is tunable (nominally from 0 to 0.4). The magnetism of our a-few-SL MBT films were investigated by Hall measurements. We adopt a sandwich structure of Bi$_2$Te$_3$ (BT)/n-SL MBT/BT, where the bottom BT layer serves as a better substrate to improve the crystallinity of MBT and the top BT layer acts as a protection layer. This kind of structure is stable as shown in the STEM image of the BT/MBT/BT ($x$ = 0.3) film, which was kept in ambient conditions for 2 weeks before the STEM measurement.

Figure 1(b) shows the longitudinal resistance $R_{xx}$ versus temperature for 1-SL and 2-SL MBT films with different $Bi_{Mn}$ doping $x$ on sapphire substrates. The data displays non-monotonic behaviors with doping as shown in the 1-SL case. The resistance decreases first ($x = 0.1$) and then increases ($x = 0.28$) compared with that at $x = 0$. We attribute the decrease at lower doping to the carrier-doping effect (see Fig. S1 of [18]) and the increase to the disorder effect. The non-monotonic doping dependent resistance is basically resulted from the competing behaviors between these two effects.

*The FM and the SG-like states in 1-SL MBT*—The 1-SL MBT is suggested to be FM with a transition temperature $T_c$ of about 12 K[6,19]. In our Hall data of the $x = 0$ film in Fig. 2(a), $T_c$ is about 8 K. Above $T_c$ there are weak SG-like behaviors (in the ellipse), where $R_{xy}$ (proportional to the out-of-plane magnetic moments) shows hysteresis even in the absence of zero-field magnetism. The existence of SG-like magnetism is also proved by the ramping-rate dependent Hall loop (Fig. S3 of [18]). We propose that the reduced $T_c$ in our epitaxial 1-SL film may be resulted from the inevitable presence of $Bi_{Mn}$ defects in the Mn-layer even in the nominal $x = 0$ case. The $Bi_{Mn}$ doping in the Mn-layer breaks the condition that the nearest-neighbor (NN) exchange interactions be FM between the $Mn_{Mn}$ moments[20], where the competition between FM and AFM interactions may give rise to the SG-like behaviors. In the presence of heavy non-magnetic $Bi_{Mn}$ doping ($x = 0.28$), $T_c$ drops to about 6 K due to the reduced NN FM interactions (Fig. 1e). In addition, the SG-like behavior becomes more prominent in this case, where there is a large hysteresis in the $R_{xy}$ loop even without the zero-field magnetism at 6 K. Furthermore, $R_{xx}$ of samples with $x = 0.28$ are different from those with $x = 0.28$, indicative of different carrier-spin dynamics at 4.2 K in samples with more prominent SG-like magnetism (Fig. S4 of [18]).

For the dilute $Mn_{Bi}$ moments, it is suggested that the $Mn_{Bi}$ moments and those of adjacent $Mn_{Mn}$ ones can be considered to always be in the AFM alignment as resulted from the strong exchange interaction between them[7,21]. Normally the NN exchange interactions in the dilute $Mn_{Bi}$ case is in the FM region resulted from carrier-mediated Ruderman-Kittel-Kasuya-Yosida (RKKY) interactions. Thus, in the $x = 0$ case, the two dilute doped Bi-layers and the Mn-layer in 1-SL MBT constitute a ferrimagnetic (FiM) ground state.

*The re-entrant FM in 2-SL MBT*—In 2-SL MBT films, abnormal magnetic behaviors appear as shown in the anomalous Hall (AH) data in Figs. 2(c)-2(g) for films with different $x$. In the $x = 0$ case, there appears FM-like behavior at the lowest temperature (0.4 K), which weakens with the increasing temperature and almost disappears around 6 K ($T_{c1}$). Further increasing the temperature leads to the appearance of another FM-like state with a reversed Hall polarity and with different cohesive fields. The FM state in the higher temperature region disappears at around 14.5 K ($T_{c2}$). The dashed curves indicate the spin-flop fields at different temperatures for the alignment of inter-SL $Mn_{Mn}$ moments, indicating the AFM inter-SL alignment among majority $Mn_{Mn}$ moments. In addition, the simultaneously measured $R_{xx}$ signals show huge discrepancy between forward and backward sweeping curves, which can be relaxed in the time scale of several minutes (Fig. 2(e) or Fig. S5 of [18]). This signifies the enhancement of competing interactions in the 2-SL case compared with the 1-SL one, which is reasonable in that the dilute FM becomes stronger with increased number of Bi-layers and that there appear additional AFM interactions between the two Mn-layers.

This kind of re-entrant FM behavior disappears at the intermediate doping level of $x = 0.1$ (Fig. 2(f)). By the comparison between the line shapes of the AH loops of the $x = 0$ and $x = 0.1$ films, we see that the disappearance of the inverted FM-like behavior at low temperatures comes from the appearance of the dip-like features just below the spin-flop fields, which shift the forward and backward curves in opposite directions and annihilate the crossings in the $x = 0$ case at low temperatures. The FM $T_{c2}$ barely changes at $x = 0.1$ but drops to around 6 K by further increasing the $Bi_{Mn}$ doping to $x = 0.28$ as shown in Fig. 2(g).

*The symmetric exchange-bias effect*—We attribute this kind of re-entrant FM behavior to a kind of symmetric exchange-bias effect as proposed below that is different from the normal ones observed in single crystals or heterostructures [22-26]. In 2-SL MBT films, we divide the inter-SL NN $Mn_{Mn}$ pairs into three situations. Types A and C denote the $Mn_{Mn}$ pairs with and without a NN $Mn_{Bi}$ moment as shown in Fig. 3(a), respectively. Type B also denotes the $Mn_{Mn}$ pairs with a NN $Mn_{Bi}$ moment but with additional NN $Bi_{Mn}$ defects, which are nonmagnetic and reduce the FM coupling between the type-B $Mn_{Mn}$ moments and the nearby ones. Because each Bi-layer has a $Mn_{Bi}$ doping level of $y \approx 0.1$, the proportion of type-A and -B moments sum up to 0.4. Note that in nominally $x = 0$ samples, type-B pairs are suppressed. Here the type-C moments response to the field as the MBT without anti-site defects, aligning anti-ferromagnetically and ferromagnetically below and above the spin-flop fields $H_{sf}$, respectively. The peculiar magnetic behaviors lie in the competing exchange interactions $J_1^+, J_2^-, J_3^-$ and $J_4^+$ as shown in Fig. 3(a), which correspond to the NN FM, AFM, FM and AFM interactions between intra-SL $Mn_{Mn}$ moments, intra-SL $Mn_{Bi}$ and $Mn_{Mn}$ moments, inter-SL $Mn_{Mn}$ moments, and inter-SL $Mn_{Bi}$ moments, respectively. Here the superscript denotes the sign of the exchange interaction.

Type A moments can be further evenly divided into two sub-types A1 and A2 depending on which SL the Mn$_{Bi}$ moment lies in, as shown in the solid- (A1) and dashed-squares (A2) in Fig. 3(b). Considering the situation that after sweeping the field downward across $H_{sf}$ type-C moments are in the ↓↑ configuration (the second step in Fig. 3(a)), the Hamiltonian of exchange interactions in the type-A2 case $-J_1^+ \sum_{NN} \boldsymbol{M}_1 \boldsymbol{M}_{1i} - J_1^+ \sum_{NN} \boldsymbol{M}_2 \boldsymbol{M}_{2i} - J_3^- \boldsymbol{M}_1 \boldsymbol{M}_2 - J_2^- \boldsymbol{M}_2 \boldsymbol{m}_2 - J_4^+ \sum \boldsymbol{m}_2 \boldsymbol{m}_{2i} - \mu_0 H(\boldsymbol{M}_1 + \boldsymbol{M}_2 + \boldsymbol{m}_2)$ becomes $-N_{\boldsymbol{M}_1} J_1^+ \boldsymbol{M}_1 \downarrow - N_{\boldsymbol{M}_2} J_1^+ \boldsymbol{M}_2 \uparrow - J_3^- \boldsymbol{M}_1 \boldsymbol{M}_2 - J_2^- \boldsymbol{M}_2 \boldsymbol{m}_2 - J_4^+ \sum \boldsymbol{m}_2 \boldsymbol{m}_{2i} - \mu_0 H \boldsymbol{M}_1$, where $\boldsymbol{M}_1, \boldsymbol{M}_2, \boldsymbol{M}_{1i}, \boldsymbol{M}_{2i}, \boldsymbol{m}_1, \boldsymbol{m}_{1i}, N_{\boldsymbol{M}_1}, N_{\boldsymbol{M}_2}$ are the Mn$_{Mn}$ moments in the top and bottom SLs, the NN-Mn$_{Mn}$ moments of $\boldsymbol{M}_1$ and $\boldsymbol{M}_2$, the Mn$_{Bi}$ moment and the Mn$_{Bi}$ moments near $\boldsymbol{m}_1$, the numbers of the NN Mn$_{Mn}$ moments for $\boldsymbol{M}_1$ and $\boldsymbol{M}_2$, respectively. Note that $\boldsymbol{M}_2$ and $\boldsymbol{m}_2$ are always in the AFM alignment. In addition, we assume the same magnitudes of Mn$_{Mn}$ and Mn$_{Bi}$ moments so that the total Zeeman energies of $\boldsymbol{M}_2$ and $\boldsymbol{m}_2$ is zero. The energy difference between the ↓↓↑ and ↑↓↑ configurations of type-A2 moments is then $-2N_{\boldsymbol{M}_1} J_1^+ + 2\mu_0 H$, the critical condition of which is the same with the type-C case and determines the spin-flop field. Here the bigger and smaller arrows denote the Mn$_{Mn}$ and Mn$_{Bi}$ moments, respectively. Thus, between steps 1 and 2 the type-A2 sub system undergoes spin-flop transition similarly to type-C moments, keeping the FM Mn$_{Bi}$ moments unchanged.

For the type-A1 case, considering an area $S$ with a characteristic length scale $\propto 1/k_F$ of RKKY interactions in dilute-doped Bi-layers, the energy difference between the magnetic alignments that all the type-A1 moments are in the same ↑↓↑ or ↓↑↑ configurations is $-2S n_{A1} N_{\boldsymbol{M}_1} J_1^+ + S^2 n_{A1} n_{A2} J_4^+$, where $n_{A1} \approx n_{A2} = n$ is the density of types-A1 and -A2 moments, respectively. In our MBT films with $y \approx 0.1$, $n \approx 1.2$ /nm$^2$. Thus. there is a critical doping level $y$ above which all the dilute Mn$_{Bi}$ moments align ferromagnetically. We propose that the doping level of 0.1 in our films is above such criticality. Thus, from step 1 to step 2, all the Mn$_{Bi}$ and type-A1 Mn$_{Mn}$ moments keep the same alignment with that above $H_{sf}$.

At step 3, with the direction of field reversed, it's easy to see that the type-A2 Mn$_{Mn}$ moments are biased instead compared with step 2 and the Mn$_{Bi}$ moments flips to the opposite direction. Similar analysis can be taken on the backward field sweep. This leads to the AH loop illustrated in Fig. 3(c). We call this behavior a kind of symmetric exchange-bias effect, where the net magnetization of the AFM system is exchange-biased by the dilute FM system and responses to the external field with a cohesive field $\mu_0 H_{eb} \sim 0.27$ T in a symmetric way. The change in the Hall signal from step 2 (↑↑↓↓) to step 3 (↓↓↑↑) may come from the different coefficients that Mn$_{Mn}$ and Mn$_{Bi}$ moments interact with the carriers. The dashed lines indicating the situation without the exchange-bias effect, which is proposed for even-layer MBT films without anti-site defects.

Figure 4(a) summarizes the magnetic configurations at each step for both type-A and type-C moments that dominate the $x = 0$ sample. We see that the Hall signals of step 2 and step 3 in different sweeping directions don't coincide at low temperatures. This comes from the SG-like nature of the system, which is driven into metastable states after the spin-flop transition and relaxes slowly with time. When sweeping the field below $H_{sf}$, this mismatch diminishes as shown in Fig. 2(d). With the increasing temperatures, the intra-Mn-layer exchange interaction $J_1^+$ decreases faster than the dilute one ($J_4^+$) as suggested in our other work [in preparation]. This proposition is also supported by the fact that the higher-temperature FM starts to appear above $T_c$ of 1-SL MBT, which is purely decided by $J_1^+$. Hence, in the higher temperature region, the type-A moments response to the field similarly to type-B ones as shown in Fig. 4(b). The inter-SL Mn$_{Mn}$ moments still keep the AFM alignment at low fields as can be seen from the nearly unchanged signal step caused by the spin-flop transition. The intra-SL Mn$_{Mn}$ alignment becomes disturbed instead, keeping a FM alignment among the Mn$_{Bi}$ moments. The magnetic configurations at each step in this case is shown along with the AH loop at 12.0 K.

*The disappearance of re-entrant FM with Bi$_{Mn}$ doping*—The disappearance of the re-entrant phenomenon at $x = 0.1$ is proposed to be similar to the $x = 0$ case above $T_c$. The nonmagnetic doping of Bi$_{Mn}$ into the Mn-layers reduces the intra-SL FM interaction, introducing the type-B moments which response to the external field as mentioned above even at low temperatures. As shown in Fig. 4(c) or Fig. 2(f), the coexistence of type-A and type-B moments leads to the AH loops different from those of the $x = 0$ film. Compared with the AH loop of the $x = 0$ sample, we see that the existence of type-B components pushes the forward and backward curves in opposite directions and annihilates the inverted FM state, though the exchange-bias step is still there between steps 2 and 3. The different coercive fields for the type-B moments and the exchange-biased type-A moments introduce an additional step in the original step 3 in the AH curve.

*Conclusion*—We successfully prepared 1- and 2-SL MBT epitaxial films with controlled Bi$_{Mn}$ doping in the Mn-layers. We show that when two FiM 1-SL MBT are stacked with each other, the three competing interactions (the inter-SL AFM interactions between Mn$_{Mn}$ moments in Mn-layers, the FM interactions among all the Mn$_{Bi}$ moments, and the strong AFM interactions between these two sub-systems) lead to a kind of symmetric exchange-bias effect that shows as the re-entrant FM behavior. This re-entrant behavior disappears in films with a Bi$_{Mn}$-doping level of about 0.1. Our work achieves the fine-tuning of anti-site defects in the ultra-thin MBT epitaxial films and clarifies the complex doping-dependent magnetism, which is crucial to the realization of exotic states promised for this kind of magnetic topological insulator.

**Acknowledgments**

We acknowledge the supporting from National Key R&D Program of China (Grants No. 2022YFA1403102) and National Science Foundation (Grants No. 92065102, 61804056, 12134008).

**Author contributions**

Y-. P. J. conceived and designed the experiments. D. S., F. H., G. Y., H. Z., and H. H. carried out MBE growth and STM measurements. Y-. P. J. did the transport measurement. Y-. P. J., J. Z., and D. S. did data analyses and interpretations. L. G., Q. Z., X-. C. M., and H. Y. carried out the TEM measurements. Y-. P. J wrote the manuscript, with input from all authors.

**Competing interests**

The authors declare that they have no competing interests.

**Data and materials availability**

All data are available in the main text or the supplementary materials.

**Supplementary Materials**

Figs. S1 to S5

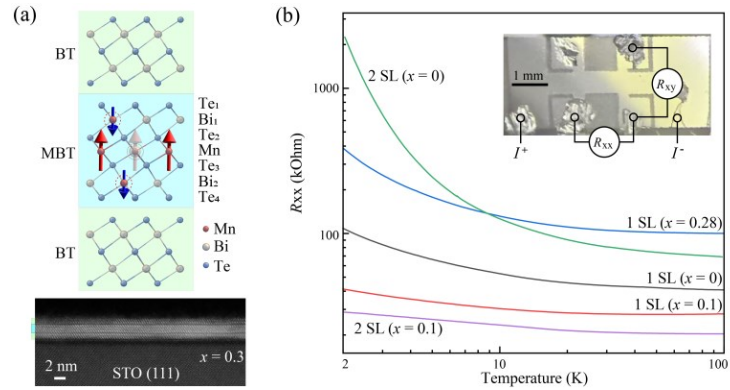

Fig. 1. (Color online) The structure and the resistance measurement of MBT films. (a) The crystal structure (top) and the cross-sectional HAADF-STEM image (bottom) of the BT/MBT/BT film ($x = 0.3$) on the STO (111) substrate. The dashed and solid circles indicate the positions of $Mn_{Bi}$ and $Bi_{Mn}$ anti-site defects, respectively. (b) The longitudinal resistances of the Hall bars for 1-SL and 2-SL MBT with different $Bi_{Mn}$ doping on the sapphire substrates. The inset is a photo of Hall bar.

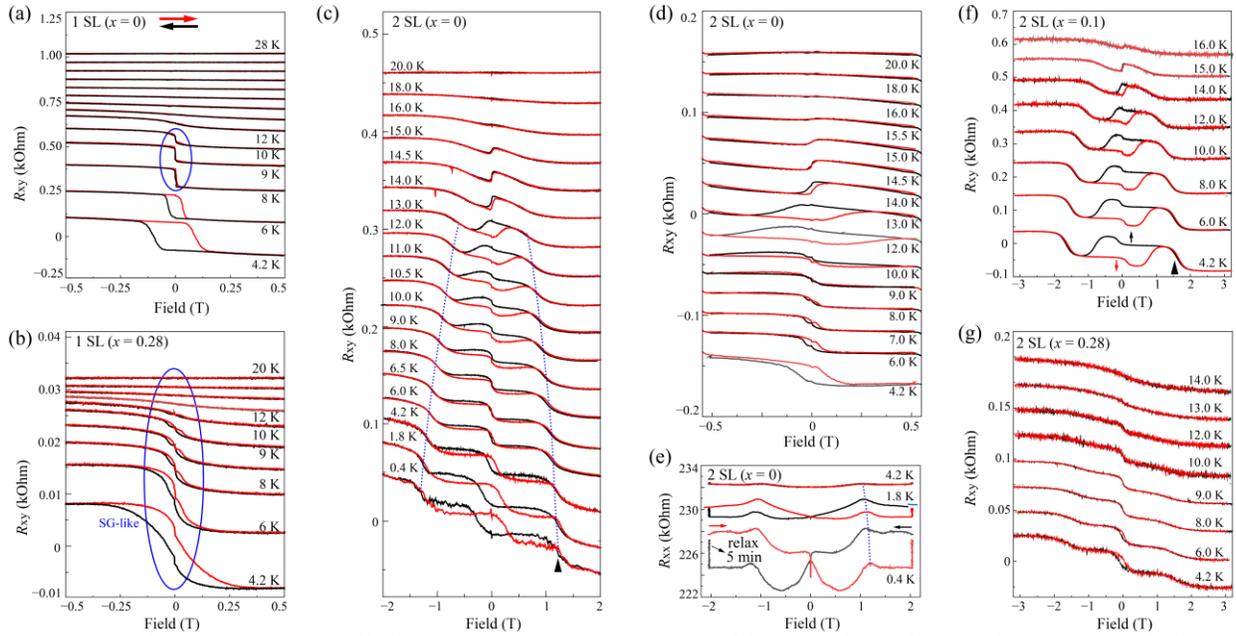

Fig. 2. (Color online) The different magnetism in MBT films with different doping and different thicknesses. (a)(b) The temperature-dependent Hall resistances for 1-SL MBT with $x = 0$ and $x = 0.3$, respectively. The arrows indicate the sweeping directions of the magnetic field. (c)(d) The temperature-dependent AH data with the field swept in different ranges for 2-SL MBT ($x = 0$). (e) The longitudinal resistance versus field. The dashed curves indicates the fields for spin-flop transition at different temperatures. (f)(g) The temperature-dependent AH data for 2-SL MBT with $x = 0.1$ and $x = 0.28$, respectively.

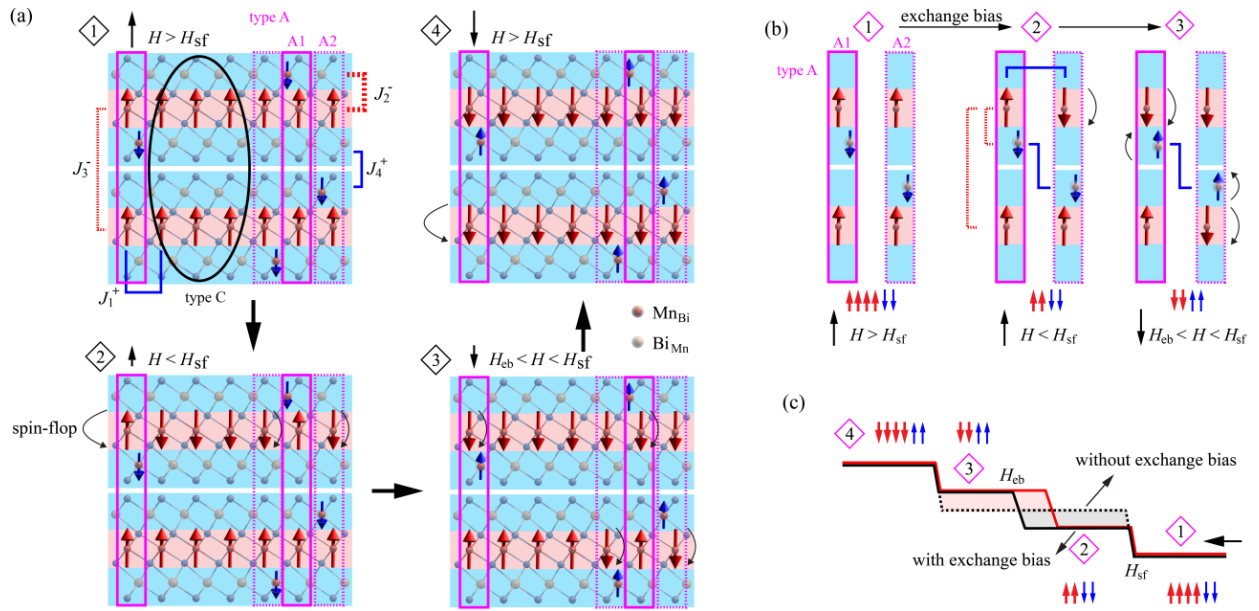

Fig. 3. (Color online) The origin of the exchange-bias effect responsible for the re-entrent FM in 2-SL MBT. (a) The magnetic alignments of the spins in 2-SL MBT ($x = 0$) at different stages during the field sweep. The solid and dashed rectangles as well as the ellipse denote type-A1, -A2, and -C blocks, recpectively. (b) The detailed behaviors of types A1 and A2 in different fields. (c) The different shapes of the AH loops w/o the exchange-bias effect.

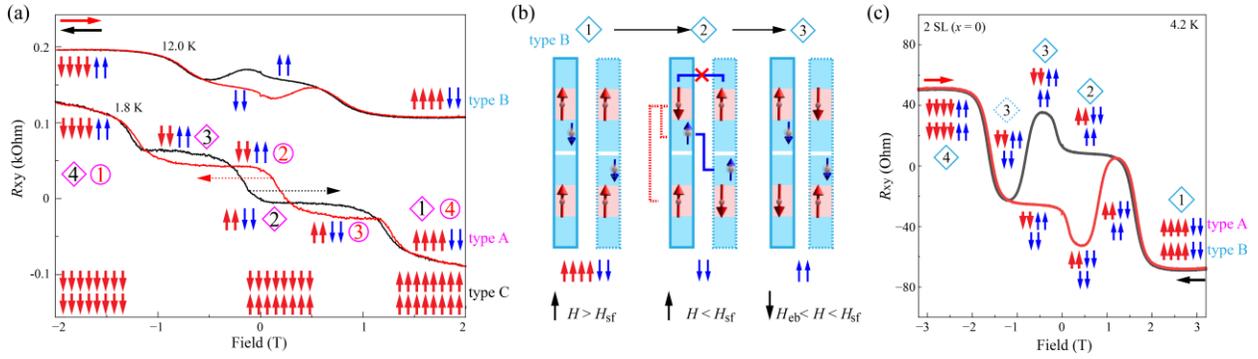

Fig. 4. (Color online) (a) The evolvement of magnetic alignments for two different FM states at 1.8 K and 12.0 K in Fig. 2(c). The dashed arrows indicate the trend of change of the AH loop with the increasing temperature. (b) The detailed alignments of the type-B moments in different fields in the absence of $J_1^+$. (c) The evolvement of magnetic alignments in the presence of both type-A and type-B magnetic blocks for 2-SL MBT ($x$ = 0.1) at 4.2 K in Fig. 2(f).

# Supplementary Materials for

## The exchange-bias effect and the re-entrant ferromagnetism in MnBi$_2$Te$_4$

Dezhi Song[1], Fuyang Huang[1], Gang Yao[2], Haimin Zhang[1], Haiming Huang[1], Hang Yan[3], Jun Zhang[1*], Qinghua Zhang[4], Lin Gu[5], Xu-Cun Ma[3,6], Jin-Feng Jia[2,7,8], Qi-Kun Xue[4,6,7,8], Ye-Ping Jiang[1*]

1 Key Laboratory of Polar Materials and Devices, Department of Electronic, East China Normal University, Shanghai 200241, China.

2 Key Laboratory of Artificial Structures and Quantum Control (Ministry of Education), Tsung-Dao Lee Institute, School of Physics and Astronomy, Shanghai Jiao Tong University, Shanghai, 200240, China.

3 State Key Laboratory of Low-Dimensional Quantum Physics, Department of Physics, Tsinghua University, Beijing 100084, China.

4 Beijing National Laboratory for Condensed Matter Physics, Institute of Physics, Chinese Academy of Sciences, Beijing 100190, China.

5 Beijing National Center for Electron Microscopy and Laboratory of Advanced Materials, School of Materials Science and Engineering, Tsinghua University, Beijing 100084, China.

6 Frontier Science Center for Quantum Information, Beijing 100084, China.

7 Southern University of Science and Technology, Shenzhen 518055, China.

8 Quantum Science Center of Guangdong-HongKong-Macao Greater Bay Area, Shenzhen 518045, China.

*Corresponding author. Email: zhangjun@ee.ecnu.edu.cn, ypjiang@clpm.ecnu.edu.cn

**The PDF file includes:**
Supplementary Sections I-II
Supplementary Figs. S1 to S5

**I. Sample preparation**

To achieve precise control of anti-site defects in MBT, we first do the MBE growth and in-situ STM characterization on the conductive STO (111) substrates. The STO substrates were annealed up to 800 °C to get an atomically flat surface prior to the deposition of BT and MBT. The first QL of BT grown on STO acts as an ideal epitaxial substrate for the growth of 1-SL MBT. All the heterostructures of MBT and BT were grown in a step-wise manner, that is, only one QL or SL was deposited and annealed at 250 °C each time. The precise control of the nominal $Bi_{Mn}$ doping was achieved by fine control of the flux ratio between Bi and Mn by using a quartz crystal monitor. The quality of the film was checked by STM and Rheed (Fig. S1 and Fig. S2) between successive growth steps. The STO substrate has atomically smooth terraces.

In the inevitable presence of Mn-Bi anti-site defects, MBT is of the form $Mn_{1-x+2y}Bi_{2(1-y)+x}Te_4$. We find the dilute doping of Mn in the Bi-layer ($Mn_{Bi}$) is nearly constant ($y \sim 0.1$) in our films. The $Mn_{Bi}$ doping is measured by counting the defects in the STM image (Fig. S1(b)). On the contrary, the Bi doping $x$ in the Mn-layer can be tuned. We successfully prepared 1-SL MBT with $x$ ranging from 0 to 0.4. The x doping condition can be double-checked by the resistance-switch behaviors (described in our other work) which are sensitive to the $Bi_{Mn}$ defects. The films keep good crystallinity as shown in the TEM image for the $x = 0.3$ one (Fig. 1).

**II. Transport measurement**

For transport measurements, the BT/MBT/BT and BT/MBT/MBT/BT samples of different doping were grown on sapphire substrates with the same growth recipes of those on STO. The sapphire substrates were annealed in a furnace at 1000 °C for one hour with an oxygen flow to get step-grade quality and then annealed in vacuum at 600 °C for half an hour prior to the growth of films. The samples were etched mechanically into Hall bars in air within one hour and transferred back into the UHV chamber. The Hall measurements were carried out in our UHV chamber by using home-built electronics. The alternating currents (1 μA, 19.875 Hz) were output from a Stanford SR830 and the $R_{xx}$, $R_{xy}$ signals were measured by two SR830 simultaneously. The temperature dependent $R_{xx}$ data was measured by PPMS.

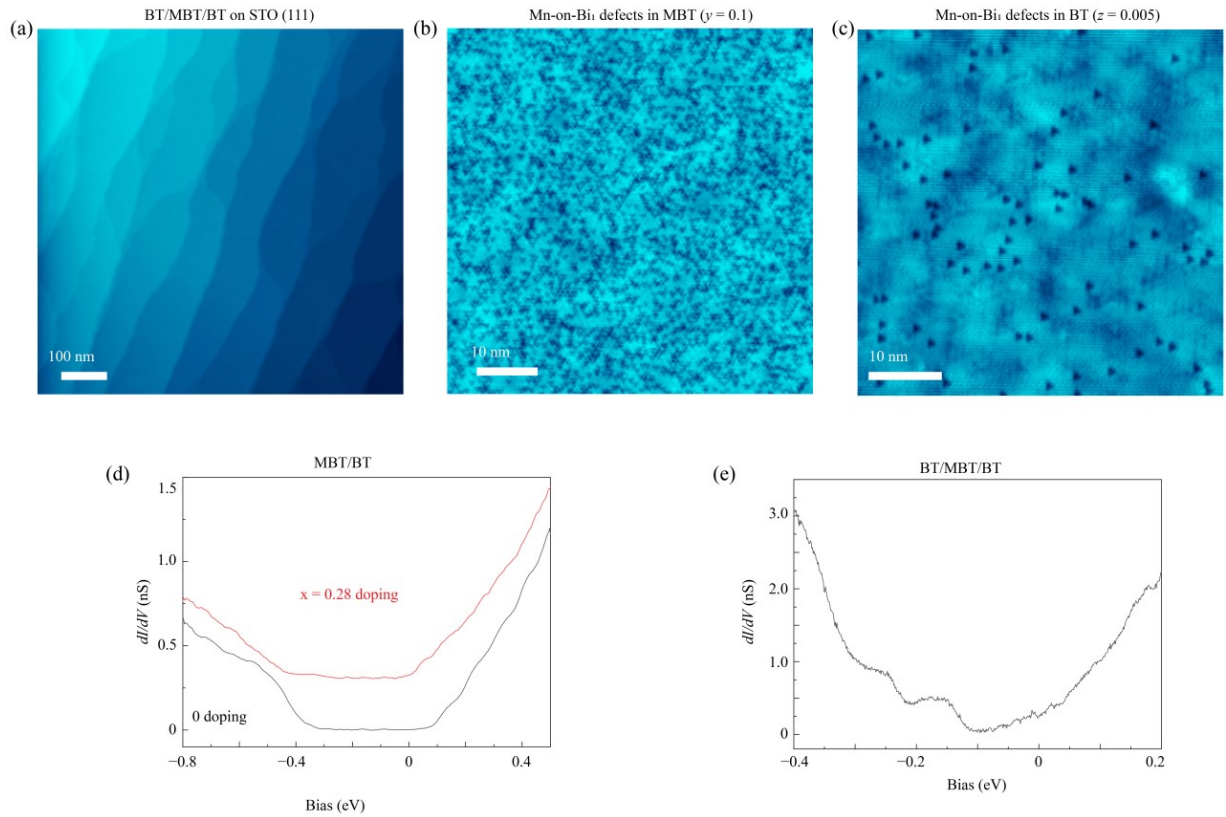

Fig. S1. The BT/MBT/BT film grown with the same procedure on STO (111). (a) The film morphology of the BT/MBT/BT film. (b)(c) The STM images showing the density of the $Mn_{Bi}$ defects in MBT (measured on the MBT/BT film) and BT (measured on the BT/MBT/BT film), respectively. (d) The dI/dV spectra taken on MBT/BT films with different $Bi_{Mn}$ doping. (e) The spectrum taken on the BT/MBT/BT film.

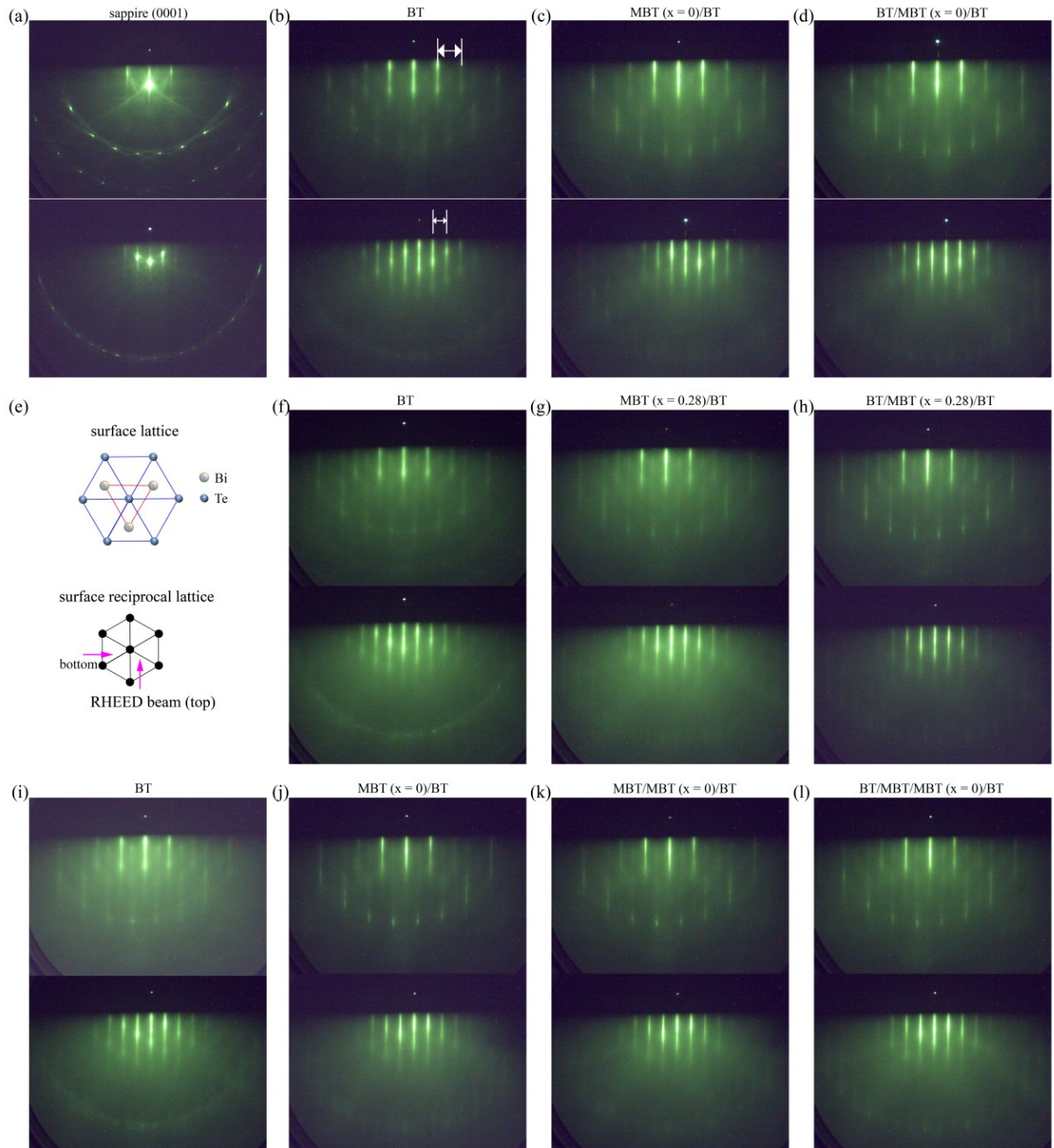

Fig. S2. (Color online) The RHEED patterns for films at different growth stages. (a) The RHEED patterns for the sapphire substrate with different directions of incident electron beam as illustrated in (e). (b)-(d) The RHEED patterns for the BT/MBT ($x = 0$) /BT film at different growth steps. (e) The surface lattice of BT or MBT and the two directions of incident beam with respect to the surface reciprocal lattice. (f)-(h) The RHEED patterns for the BT/MBT ($x = 0.28$) /BT film at different growth steps. (i)-(l) The RHEED patterns for the BT/MBT/MBT ($x = 0$) /BT film at different growth steps.

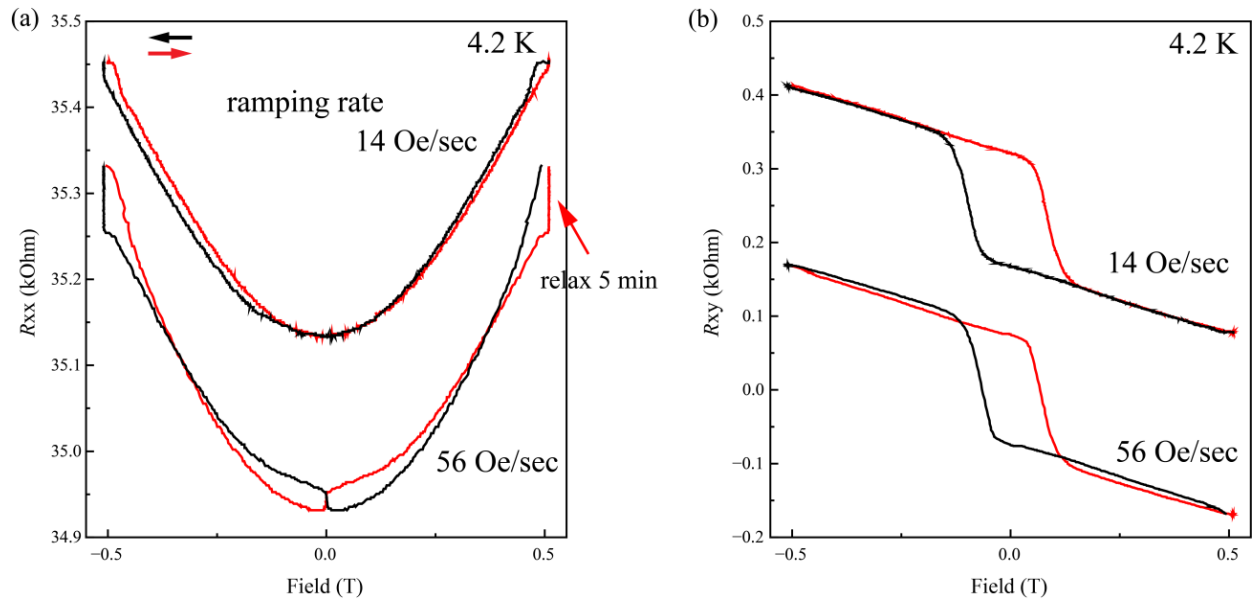

Fig. S3. (Color online) (a)(b) The $R_{xx}$ and $R_{xy}$ loops for 1-SL MBT ($x = 0$) with two different ramping rate of field at 4.2 K, respectively.

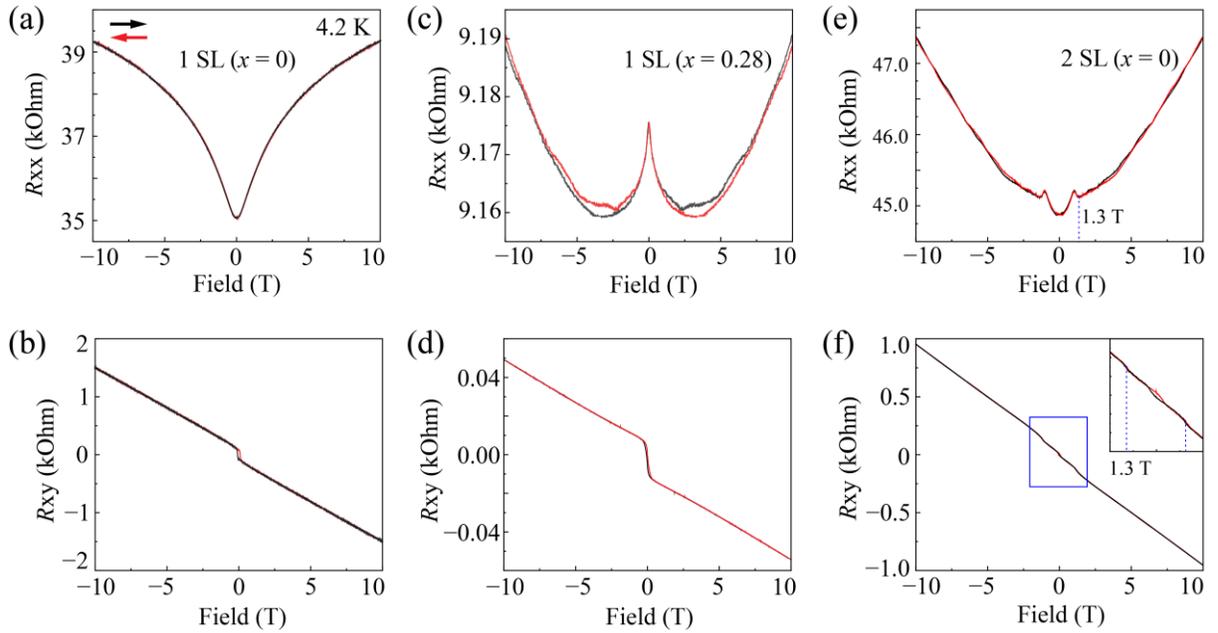

Fig. S4. (Color online) (a)-(f) The $R_{xx}$ (top) and $R_{xy}$ (bottom) loops for 1-SL MBT ($x = 0$), 1-SL MBT ($x = 0.28$) and 2-SL MBT ($x = 0$) films in the field range of ± 10 T, respectively.

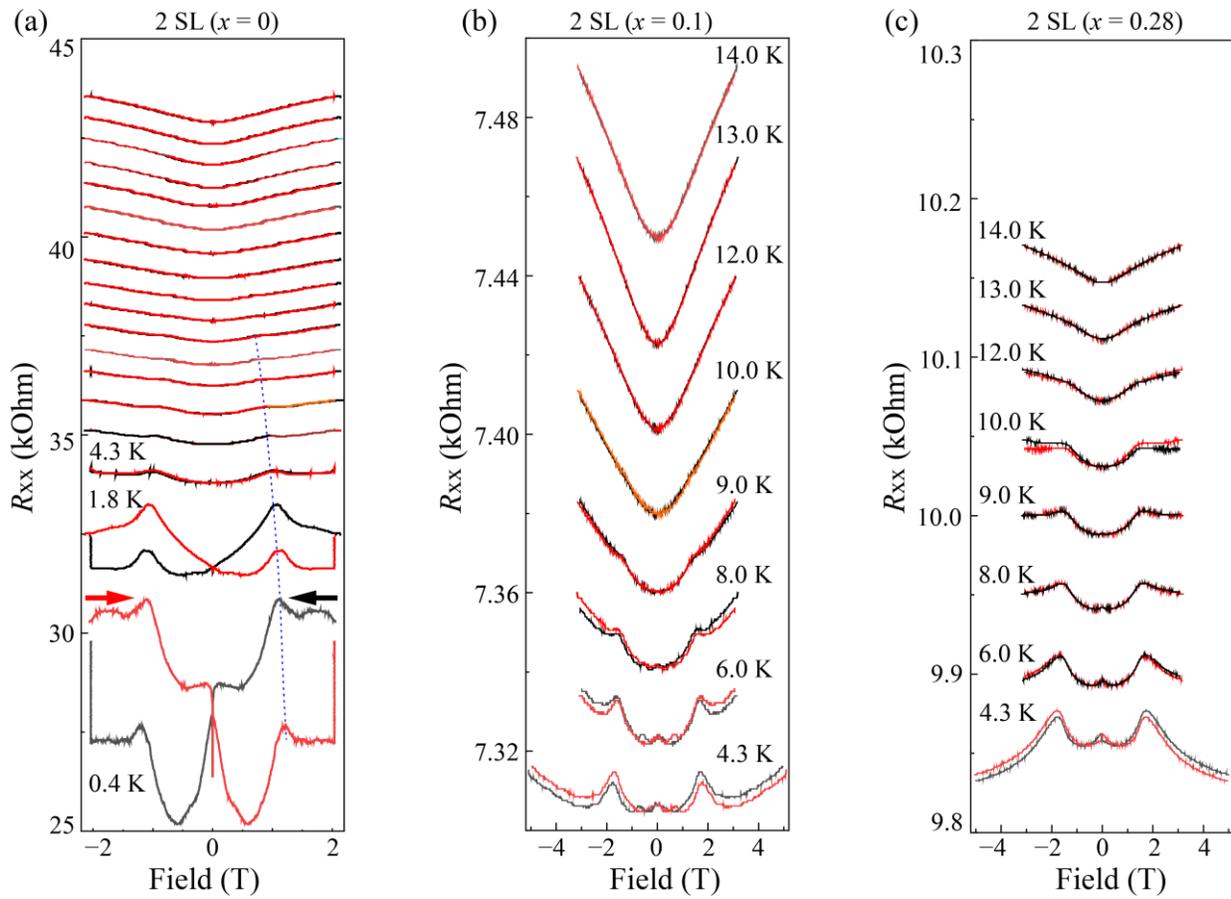

Fig. S5. (Color online) (a)-(c) The $R_{xx}$ loops with the field swept between ± 5 T for 2-SL MBT films with different doping $x$ at different temperatures.